\documentclass[aps,onecolumn,groupedaddress,amsmath,amssymb,nofootinbib,superscriptaddress]{revtex4-2}
\usepackage{amsmath, dsfont}
\usepackage{amssymb}
\usepackage{epsfig}
\usepackage{calc}
\usepackage{amsfonts}
\usepackage{enumerate}
\usepackage{graphicx}
\usepackage[ansinew]{inputenc}
\usepackage{multirow}
\usepackage{subcaption}
\usepackage{tensor}

\usepackage[dvipsnames]{xcolor}
\usepackage{natbib}
\usepackage{hyperref}
\usepackage{comment}
\usepackage{soul}

\usepackage{times}

\graphicspath{{Figs/}}

\definecolor{azure(colorwheel)}{rgb}{0.0, 0.5, 1.0}
\definecolor{PRDblue}{RGB}{48,46,146}
\definecolor{nicered}{rgb}{0.7,0.1,0.1}
\definecolor{DarkViolet}{RGB}{148,0,211}

\hypersetup{colorlinks,bookmarksopen,
bookmarksnumbered,
linkcolor=PRDblue, 
filecolor=PRDblue,
urlcolor=teal,
citecolor=nicered
}

\newcommand{\G}{{\mathcal{G}}}

\usepackage[normalem]{ulem}

\usepackage{orcidlink}
\def\idbako{\orcidlink{0000-0002-3012-6144}}
\def\idnaka{\orcidlink{0000-0002-3522-5803}}
\def\idchar{\orcidlink{0000-0002-5364-4753}}
\def\idkand{\orcidlink{0000-0002-3018-5558}}
\def\idleco{\orcidlink{0000-0001-8686-4093}}

\begin{document}
\setcounter{page}{1}
\title[]{Black holes with primary scalar hair}
\author{Athanasios Bakopoulos\idbako}
\affiliation{Physics Division, School of Applied Mathematical and Physical Sciences, National Technical University of Athens, Zografou Campus, Zografou, GR-15780, Greece}
\affiliation{Division of Applied Analysis, Department of Mathematics, University of Patras, Rio Patras GR-26504, Greece}
\author{Christos Charmousis\idchar}
\affiliation{Universit\'e Paris-Saclay, CNRS/IN2P3, IJCLab, 91405 Orsay, France}
\author{Panagiota Kanti\idkand}
\affiliation{Division of Theoretical Physics, Physics Department, University of Ioannina, GR 45110 Ioannina, Greece}
\author{Nicolas Lecoeur\idleco}
\affiliation{Universit\'e Paris-Saclay, CNRS/IN2P3, IJCLab, 91405 Orsay, France}
\author{Theodoros Nakas\idnaka}
\affiliation{Physics Division, School of Applied Mathematical and Physical Sciences, National Technical University of Athens, Zografou Campus, Zografou, GR-15780, Greece}

\begin{abstract}
We present explicit black holes endowed with primary scalar hair within the shift-symmetric subclass of Beyond Horndeski theories. These solutions depend, in addition to the conventional mass parameter, on a second free parameter encoding primary scalar hair. The properties and characteristics of the solutions at hand are analyzed with varying scalar charge. We observe that when the scalar hair parameter is close to zero or relatively small in comparison to the black hole mass, the solutions closely resemble the Schwarzschild spacetime. As the scalar hair increases, the metric solutions gradually depart from General Relativity. Notably, for a particular relation between mass and scalar hair, the central singularity completely disappears, resulting in the formation of regular black holes or solitons. The scalar field accompanying the solutions is always found to be regular at future or past horizon(s), defining a distinct time direction for each. As a final byproduct of our analysis, we demonstrate the existence of a stealth Schwarschild black hole in Horndeski theory with a non-trivial kinetic term. 

\end{abstract}
\maketitle
\section{Introduction}
In General Relativity (GR) coupled to electrodynamics, a stationary black hole is completely characterized by its mass $M$, angular momentum $J$ and electric charge $Q$: indeed, two such black holes with identical $M$, $J$ and $Q$ are described by the exact same Kerr-Newman metric~\cite{Kerr:1963ud}. They have \textit{no hair}, i.e., no other independent, externally observable physical quantity~\cite{Bekenstein:1971,Teitelboim:1972,Mazur:2000pn,Chrusciel:2012jk}.
Accordingly, the expression ``hairy black hole'' refers to a black hole  possessing one of the following two kinds of hair~\cite{Herdeiro:2015waa}:  \textit{primary hair}, which is a global charge distinct from mass, angular momentum or electric charge; and  \textit{secondary hair}, where the black-hole metric is dressed with non-trivial additional fields (i.e. other than electromagnetic) but remains entirely determined by $M$, $J$ and $Q$.

The quest for hairy black holes can be pursued either in GR with matter fields other than Maxwell fields, or in modified gravity theories. In the former case, hairy black holes are typically obtained by considering minimally-coupled non-abelian gauge fields~\cite{Volkov:1998cc,Karakasis:2022xzm}, skyrmions~\cite{Droz:1991cx} or scalar field (complex~\cite{Herdeiro:2015gia} or real, see e.g.~\cite{Bakopoulos:2021dry,Bakopoulos:2023hkh,Karakasis:2023hni} and references within). On the other hand, most modified gravity theories can be cast, at least in certain limits, into a scalar-tensor form, that is, a theory of gravity which includes scalar field(s) non-minimally coupled to the usual metric tensor field $g_{\mu\nu}$. This class of theories includes the superstring effective theories \cite{Callan:1986,Gross:1986mw,Metsaev:1987zx} and compactifications of higher-dimensional gravitational theories such as Lovelock theory \cite{Langlois:2018dxi}. The most general scalar-tensor theory of a single real scalar field $\phi$ leading to second-order field equations is Horndeski theory~\cite{Horndeski:1974wa}, which admits recent generalizations allowing for higher-order field equations but still propagating no ghost degree of freedom~\cite{Langlois:2018dxi}.

In the latter case and to the best of our knowledge, no explicit black holes with primary hair have been found in such scalar-tensor theories.\,\footnote{There exist interesting examples of black holes with primary hair for a minimally-coupled (two-derivative) complex scalar field~\cite{Herdeiro:2014goa,Herdeiro:2015gia}. There, the minimally-coupled scalar is part of the energy-momentum matter tensor rather than a modification of gravity. Also, a primary hair black hole was constructed in a bi-scalar extension of Horndeski theory~\cite{Charmousis:2014zaa}, but, as stated earlier, the framework of the current article is scalar-tensor theory with a single scalar field.} However,  
numerous black holes with secondary scalar hair have been constructed quite easily evading the no-hair theorem~\cite{Hui:2012qt} (see also \cite{Sotiriou:2014pfa,Babichev:2016rlq}). For explicit solutions they can be split into \textit{stealth} and \textit{non-stealth} solutions: a stealth black hole has a Ricci flat or Einstein metric, but is hairy since it is accompanied by a non-trivial scalar field; while for a non-stealth black hole, the metric is not Ricci flat or an Einstein metric. For instance, the scalar-tensor theory with action
\begin{equation}
S\left[g_{\mu\nu},\phi\right] = \frac{1}{2\kappa c}\int\mathrm{d}^4x\sqrt{-g}\Bigl\{ R+\beta G^{\mu\nu}\partial_\mu\phi\,\partial_\nu\phi\Bigr\},\label{eq:action_stealth}
\end{equation}
where $R$ is the Ricci scalar, $G_{\mu\nu}$ the Einstein tensor and $\beta$ and $\kappa=8\pi G/c^4$ coupling constants (we work from now on in units $G=c=1$), admits a stealth Schwarzschild solution dressed with a scalar field of the form
\begin{equation}
\phi = qt+\psi\left(r\right),\label{eq:lin_time}
\end{equation}
where the precise form of $\psi\left(r\right)$ can be found in~\cite{Babichev:2013cya}. The linear time-dependence of the scalar field is compatible with the staticity of the Schwarzschild metric, because action~(\ref{eq:action_stealth}) depends on the scalar field through its derivatives only (it is said to be shift-symmetric, which means symmetric under shifts $\phi\to\phi+\text{const.}$). 
Furthermore, linear time-dependence renders the scalar field regular at the horizon and evades the no-hair theorem~\cite{Hui:2012qt}. In~(\ref{eq:lin_time}), $q$ is a free integration constant, however, it is not referred to as a primary hair since it does not appear in the metric and does not give rise to any additional charge. As it turns out the scalar field is associated to a regular congruence of geodesics painting the GR spacetime in a stealth fashion. This is understood by noting that the kinetic term on shell is constant, $X=-\frac{1}{2}\partial_\mu\phi\,\partial^\mu\phi=q^2/2$. This construction is rather general and works out for generic parity and shift-symmetric theories~\cite{Kobayashi:2014eva}. 

Although stealth solutions are generic, finding explicit asymptotically flat non-stealth solutions turns out to be not straightforward even with secondary hair. 
One considers a very particular shift-symmetric Horndeski theory which is linked to a higher-dimensional gravity theory~\cite{Fernandes:2020nbq} where we have the following solution~\cite{Charmousis:2021npl},
\begin{gather}
\mathrm{d}s^2=-f(r)\,\mathrm{d}t^2+\frac{\mathrm{d}r^2}{f(r)}+r^2\mathrm{d}\Omega^2,\quad f\left(r\right) = 1+\frac{r^2}{2\alpha}\left(1-\sqrt{1+\frac{8\alpha M}{r^3}}\right),\\[2mm]
\phi = qt + \int\frac{\pm\sqrt{q^2r^2+f(r)}-f(r)}{rf(r)}\,\mathrm{d}r,\label{eq:phi_gb}
\end{gather}
with $\mathrm{d}\Omega^2$ the metric of the unit two-sphere, and $\alpha$ a coupling of the theory. This time, the metric is clearly different from Schwarzschild (although identical at leading order when $r\to\infty$), but the hair of the black hole is only secondary since the metric is again fully characterized by a unique integration constant, its mass $M$. Again, the free integration constant $q$ appearing in the scalar field is not primary hair, since it does not appear as an independent integration constant for the metric. Notably, the scalar field remains non-trivial for $q=0$, in which case the black hole is dressed with a purely radial scalar field~\cite{Hennigar:2020lsl} completely determined by $f(r)$. In the above non-stealth example, we no longer have constant kinetic energy, however, the integration constant $q$ only affects the scalar field and not the metric. The constant $q$ renders the scalar field regular at the event and inner future horizons\,\footnote{More precisely, the scalar field is regular at the future horizons if the $+$ sign is chosen in Eq.~(\ref{eq:phi_gb}), while it is regular at the past horizons if the $-$ sign is chosen.} whereas the solution for $q=0$ is not even defined for $r\leq r_h$.  
We thus see that scalar-tensor theories with a single scalar field with linear time-dependence allow for a number of solutions with secondary scalar hair, but crucially lack primary scalar hair. This is true albeit the fact that no no-hair theorem seems to prevent this. 

Lack of primary hair is also true for other numerical or explicit solutions involving only a radially dependent scalar and no linear time-dependence (see, for example, \cite{Kanti:1995vq,Torii:1996yi,Kanti:1996gs,Maeda:2009uy,Antoniou:2017acq,Antoniou:2017hxj,Bakopoulos:2018nui,Bakopoulos:2020dfg,Bakopoulos:2021liw,Karakasis:2021rpn,Babichev:2023dhs}). A typical example involves a linear coupling of the scalar to the Gauss-Bonnet curvature invariant where now the scalar charge is found to be fixed with the mass of the black hole so that the solution is regular at the event horizon \cite{Sotiriou:2014pfa}. We can undertake a similar construction for static solutions in Horndeski \cite{Babichev:2017guv} and beyond Horndeski theories \cite{Bakopoulos:2022csr} but again the black holes have secondary hair as the scalar charge is always fixed with respect to the black-hole mass. 
All black holes of Horndeski theories and beyond are with secondary hair independently if they are stealth, non-stealth with a time-dependent scalar or not. 

In this article, we will present two examples of primary hair black-hole solutions. They are constructed in the framework of beyond Horndeski theories~\cite{Gleyzes:2014dya}, with in addition shift symmetry under $\phi\to\phi+\text{const.}$ and parity symmetry under $\phi\to-\phi$. With these symmetries, the theories are parameterized by three arbitrary functions of the scalar-field kinetic term $X=-\frac{1}{2}\partial_\mu\phi\,\partial^\mu\phi$, called $G_2$, $G_4$ and $F_4$, and the action reads
\begin{equation}
    S\left[g_{\mu\nu},\phi\right] = \frac{1}{2\kappa } \int\mathrm{d}^4 x\sqrt{-g}\Bigl\{G_2\left(X\right)+G_4\left(X\right)R+G_{4X}\left[\left(\Box\phi\right)^2-\phi_{\mu\nu}\phi^{\mu\nu}\right]+F_4\left(X\right)\epsilon^{\mu\nu\rho\sigma}\epsilon^{\alpha\beta\gamma}_{\hspace{0.5cm}\sigma}\phi_\mu\phi_\alpha\phi_{\nu\beta}\phi_{\rho\gamma}\Bigr\}.\label{eq:action}
\end{equation}
The following notations are used for brevity: $\phi_\mu=\partial_\mu\phi$, $\phi_{\mu\nu}=\nabla_\mu\partial_\nu\phi$, and a subscript $X$ means derivation with respect to $X$. The solutions presented below describe a static, spherically symmetric spacetime,
\begin{equation}
    \mathrm{d}s^2 = -h\left(r\right)\mathrm{d}t^2+\frac{\mathrm{d}r^2}{f\left(r\right)}+r^2\mathrm{d}\Omega^2,\label{eq:metric_ansatz}
\end{equation}
while the scalar field is
\begin{equation}
    \phi = qt+\psi\left(r\right),\label{eq:scalar_ansatz}
\end{equation}
where the linear-time dependence is allowed by shift symmetry. The scalar field is dimensionless, so the dimension of $q$ is $\left(\text{length}\right)^{-1}$. Due to shift symmetry, the scalar field is determined up to an irrelevant additive constant. The main difference with the previously presented solutions~(\ref{eq:lin_time}) and~(\ref{eq:phi_gb}) is that in our construction, $q$ will be an integration constant \textit{appearing in the metric} and independent of the mass $M$ of the black hole, thus ensuring the role of  primary scalar hair.
\\

In order to focus on the solution and its properties, we directly move on to the description of the two theories and their respective black-hole solutions with primary hair: Sec.~\ref{sec:afbh} presents an asymptotically flat black hole, while Sec.~\ref{sec:bhckt} shows a black hole in a theory with a canonical kinetic term. The last section is devoted to our conclusions. The identification of these two theories among the generic action~(\ref{eq:action}) and the solutions are presented in detail in the Appendix. Note that the two solutions we focus on are homogeneous, i.e. $h\left(r\right)=f\left(r\right)$. Note also that the following solutions admit an obvious de Sitter generalization by introducing a cosmological constant $\Lambda$ in the action.

\section{Asymptotically flat black hole} \label{sec:afbh}
The first theory under consideration is parameterized by two coupling constants: $\lambda$, with dimension $\left(\text{length}\right)$ and $\eta$, with dimension $\left(\text{length}\right)^4$, while the Horndeski functionals are given by
\begin{equation}
    G_2 = -\frac{8\eta}{3\lambda^2}X^2,\quad G_4 = 1-\frac{4\eta}{3}X^2,\quad F_4 = \eta.\label{eq:th1}
\end{equation}
Notice that the theory is invariant under $\lambda\to -\lambda$; thus, and for simplicity of notation, the sign of $\lambda$ will be fixed to be positive. 
Conversely, $\eta$ can take either sign. It is straightforward\,\footnote{For more information about the derivation of the solutions see Appendix A.}  to check that this theory admits a homogeneous solution [i.e. with $h=f$ in~(\ref{eq:metric_ansatz})], with two integration constants, $M$ and $q$. The metric function $f\left(r\right)$ reads
\begin{equation}
    f\left(r\right) = 1-\frac{2M}{r}+\eta q^4\left(\frac{\pi/2-\arctan\left(r/\lambda\right)}{r/\lambda}+\frac{1}{1+\left(r/\lambda\right)^2}\right),\label{eq:bh1}
\end{equation}
while the scalar field is given by
\begin{equation}
    \phi(t,r) = qt + \psi\left(r\right),\quad [\psi'\left(r\right)]^2 = \frac{q^2}{f^2\left(r\right)}\left[1-\frac{f\left(r\right)}{1+\left(r/\lambda\right)^2}\right].\label{eq:phi1}
\end{equation}
In the above, prime stands for derivation with respect to $r$ and the kinetic term corresponding to the scalar field is of the form
\begin{equation}
    X = \frac{q^2/2}{1+\left(r/\lambda\right)^2}\,.\label{eq:X1}
\end{equation}

The solution has two {\it{independent integration constants}}: $M$, representing the Arnowitt-Deser-Misner (ADM) mass~\cite{Arnowitt:1961zz}, and the primary scalar hair, $q$, which unlike stealth solutions plays an all too important role in the metric, modifying it from its GR form. When $q=0$ the scalar hair disappears and we get back a GR solution (Schwarzschild) with a trivial scalar. When both $q$ and $M$ vanish, the solution reduces to flat spacetime. The behaviour of the metric (\ref{eq:bh1}) as $r\to\infty$ is
\begin{equation}
    f\left(r\right) = 1-\frac{2M}{r}+2 \lambda^2\frac{\eta q^4}{r^2}+\mathcal{O}\left(\frac{1}{r^4}\right),
\end{equation}
whilst the scalar asymptotes infinity at null time, $\phi=q v$, if one chooses the $+$ sign in $\psi'\left(r\right)$, and $\phi=q u$ if one chooses the $-$ sign, where $v$ and $u$ are advanced and retarded null times, respectively.
The solution is asymptotically flat with ADM mass $M$ (assumed positive from now on), and primary scalar hair $q$ which scales like electromagnetic charge in the Reissner-Nordstr\"om solution of GR. 

\begin{figure}[t]
    \centering
    \begin{subfigure}[b]{0.485\textwidth}
    \includegraphics[width=1\textwidth]{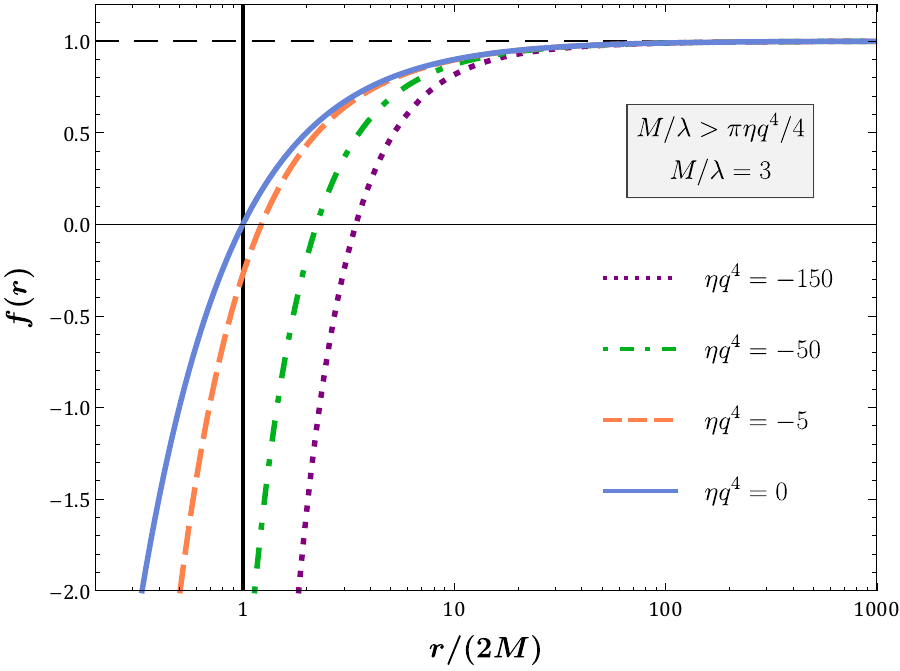}
    \caption{\hspace*{-1.5em}}
    \label{subf: sparse}
    \end{subfigure}
    \hfill
    \begin{subfigure}[b]{0.48\textwidth}
    \includegraphics[width=1\textwidth]{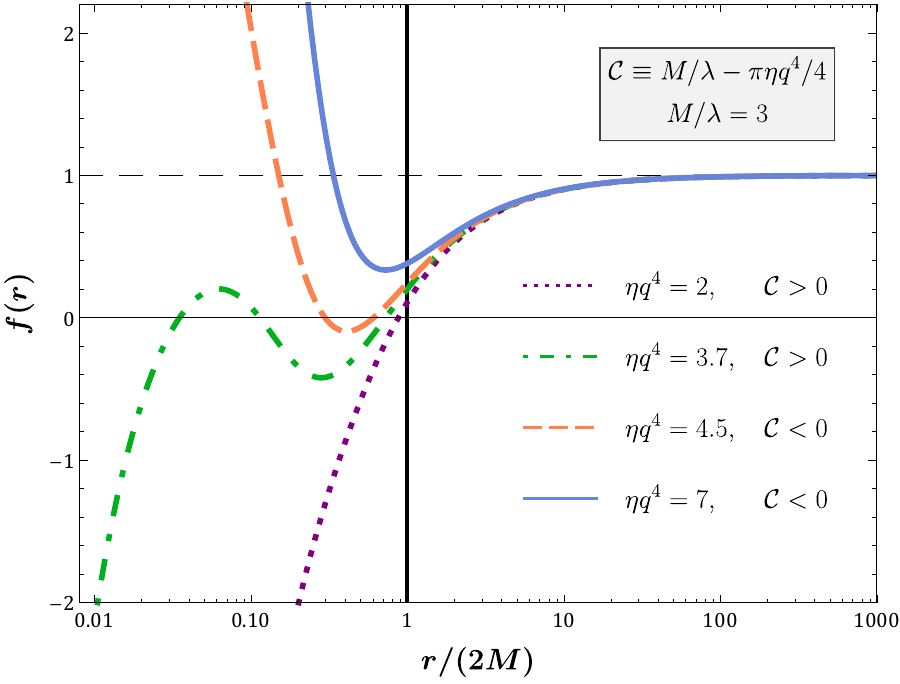}
    \caption{\hspace*{-1.5em}}
    \label{subf: ult-com}
    \end{subfigure}
    \caption{(a) $\eta<0$: single-horizon BH solutions more sparse than the Schwarzschild solution, (b) $\eta>0$: multiple-horizon BH solutions more compact than the Schwarzschild solution and a solution describing a naked singularity. The horizontal axis is logarithmic in both figures.}
    \label{fig: bh-plots}
\end{figure}

\begin{figure}[h]
    \centering
    \begin{subfigure}[b]{0.49\textwidth}
    \includegraphics[width=1\textwidth]{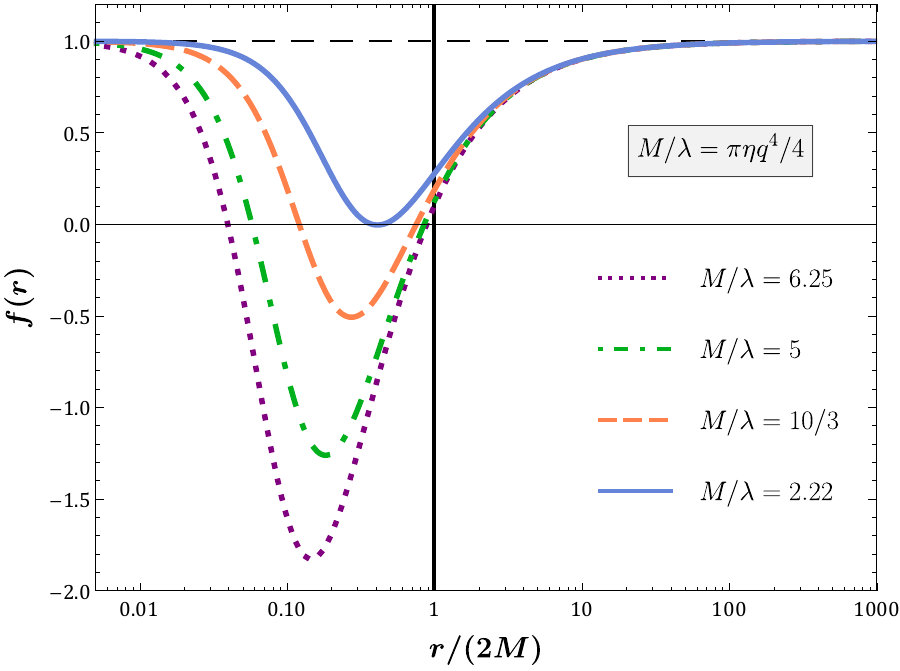}
    \caption{\hspace*{-1.5em}}
    \label{subf: rbh}
    \end{subfigure}
    \hfill
    \begin{subfigure}[b]{0.485\textwidth}
    \includegraphics[width=1\textwidth]{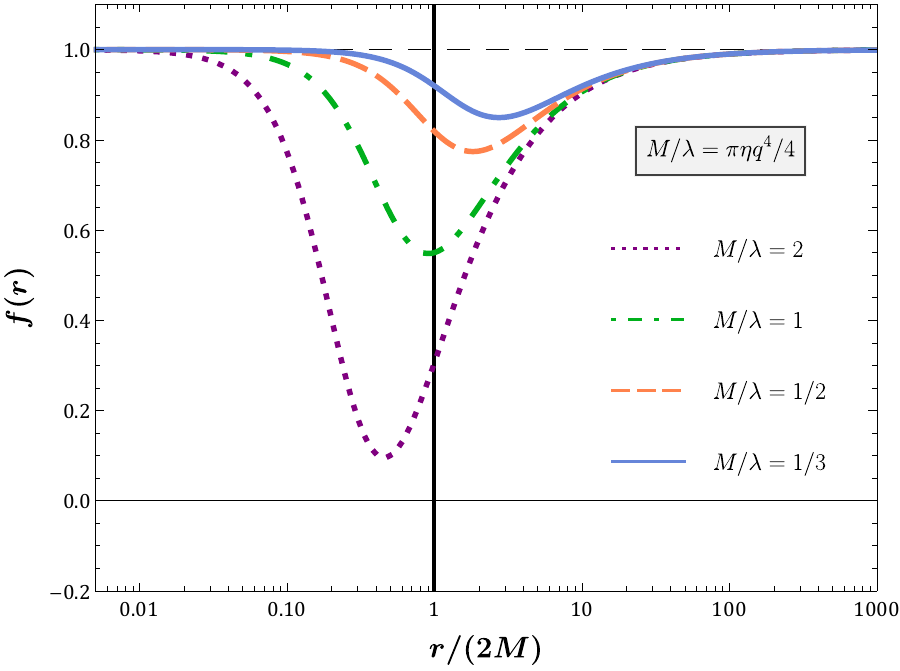}
    \caption{\hspace*{-1.5em}}
    \label{subf: sol}
    \end{subfigure}
    \caption{(a) Regular BH solutions, and (b) Solitonic solutions arising for $\eta>0$.
    The horizontal axis is logarithmic in both figures.}
    \label{fig: rbh-sol-plot}
\end{figure}

Unlike the integration constants $q$ and $M$,  the coupling constants $\lambda$ and $\eta$ fix our theory. The metric function~(\ref{eq:bh1}) shows that the independent values of $\eta$ and $q$ are not important: only the value of the product $\eta q^4$ matters. For this reason, the main characteristic of the coupling $\eta$ is its sign, since a change in its magnitude can always be accounted for by a change in the primary hair $q$. Concretely, it is possible to absorb $\eta$ in the scalar field and to end up with two distinct theories, namely, Eq.~(\ref{eq:th1}) replacing $\eta$ with $1$ or $-1$, respectively. However, we choose not to do it here in order to maintain a dimensionless scalar field.

In fact, the sign of $\eta$ plays an important role in the small-$r$ structure of spacetime. To see this, we first observe that, as $r\to 0$, we obtain
\begin{equation}
    f\left(r\right) = 1 - \frac{2M-\pi\eta q^4\lambda/2}{r}-\frac{2\eta q^4 r^2}{3\lambda^2}+\mathcal{O}\left(r^4\right). \label{eq:exp0}
\end{equation}
Then, when $\eta<0$, we clearly have $f(r)\to -\infty$, as $r\to 0$. Therefore, we always have an event horizon, with a size greater than the Schwarzschild radius $r_S=2M$, see Fig.~\ref{subf: sparse}. As the scalar charge $q$ increases the event horizon increases accordingly away from its GR size. The black hole is therefore increasingly more sparse in $q$ than its GR counterpart with $q=0$. For $\eta>0$ theories, a number of distinct geometries arise depending on the value of the mass $M$ as compared to the primary hair $q$. 
For small scalar hair, i.e. for $M/\lambda>\pi\eta q^4/4$, we have as before that $f(r)\to -\infty$ as $r\to 0$, so there is again at least one horizon. However, in this case, the event horizon size is always smaller than the Schwarzschild radius $r=r_S$, and thus the black hole is more compact compared to its GR analog, see Fig.~\ref{subf: ult-com}. As we increase the charge $q$, the event horizon shrinks in size, and three horizons emerge when $M/\lambda > \left(4+\pi\right)/4$. As the ratio $M/\lambda$ exceeds more and more this threshold,\,\footnote{The threshold value $M/\lambda=\left(4+\pi\right)/4$ is found by solving the system of equations for a triple horizon, which is then found to be located at $r=\lambda$ for the unique value $\eta q^4=2$.} the range of values of $\eta q^4$ allowing for three horizons increases. For large scalar charge, on the other hand, i.e. for $M/\lambda<\pi\eta q^4/4$, we have that $f\rightarrow +\infty$ as $r\rightarrow 0$. As the scalar charge gradually increases, we first obtain a black hole with two horizons, then an extremal solution with a double horizon, and for very large values of $q$, a naked singularity (see again Fig.~\ref{subf: ult-com}). We would also like to note here that the solution (\ref{eq:bh1}) reduces to flat spacetime only when both $M$ and $q$ vanish. If $M=0$ but $q\neq 0$, the spacetime has zero mass but is nontrivial: it is a black hole if $\eta<0$, and a naked singularity if $\eta>0$.

In the limiting case, when $M/\lambda=\pi\eta q^4/4$, we obtain classes of solutions which are completely regular. The singular term in \eqref{eq:exp0} near $r=0$ in this case disappears, and the expansion of $f\left(r\right)$ is of the form $f\left(r\right)=1+\alpha_2 r^2+\alpha_4 r^4+\alpha_6 r^6+\cdots$, containing only even powers of $r$. This is known~\cite{Burzilla:2020utr} to imply regularity of all curvature invariants and of their derivatives; that is, in addition to the regularity of the Ricci scalar $R$ or the Kretschmann scalar $K$, one has, for example, regularity of $\Box^p R$ or $\Box^p K$ for arbitrary $p$. The spacetime is parameterized by a unique integration constant, its mass $M$, which can take any positive value, and reads
\begin{gather}
    f\left(r\right) = 1-\frac{4M}{\pi\lambda}\left(\frac{\arctan\left(r/\lambda\right)}{r/\lambda}-\frac{1}{1+\left(r/\lambda\right)^2}\right),\label{eq:reg}\\[2mm]
    \phi = \left(\frac{4M}{\pi\eta\lambda}\right)^{1/4}t + \psi\left(r\right),\quad [\psi'\left(r\right)]^2 = \left(\frac{4M}{\pi\eta\lambda}\right)^{1/2}\frac{1}{f^2\left(r\right)}\left[1-\frac{f\left(r\right)}{1+\left(r/\lambda\right)^2}\right]. \label{regular}
\end{gather}
We observe that the function $f\left(r\right)$ becomes in this case an even function of $r$ thus justifying the presence of only even powers of $r$ in its expansion near $r=0$. 
In this case, we find that there exists a threshold value $a\approx 2.2116$ of the ratio $M/\lambda$ that discriminates between different types of regular solutions. Thus, the spacetime \eqref{eq:reg} describes a regular black hole with two horizons if $M/\lambda>a$, a regular extremal black hole with a double horizon if $M/\lambda=a$ (see Fig.~\ref{subf: rbh}), and a regular soliton with no horizon if $M/\lambda<a$ (see Fig.~\ref{subf: sol}). 
 
\begin{figure}[t]
    \centering
    \begin{subfigure}[b]{0.485\textwidth}
    \includegraphics[width=1\textwidth]{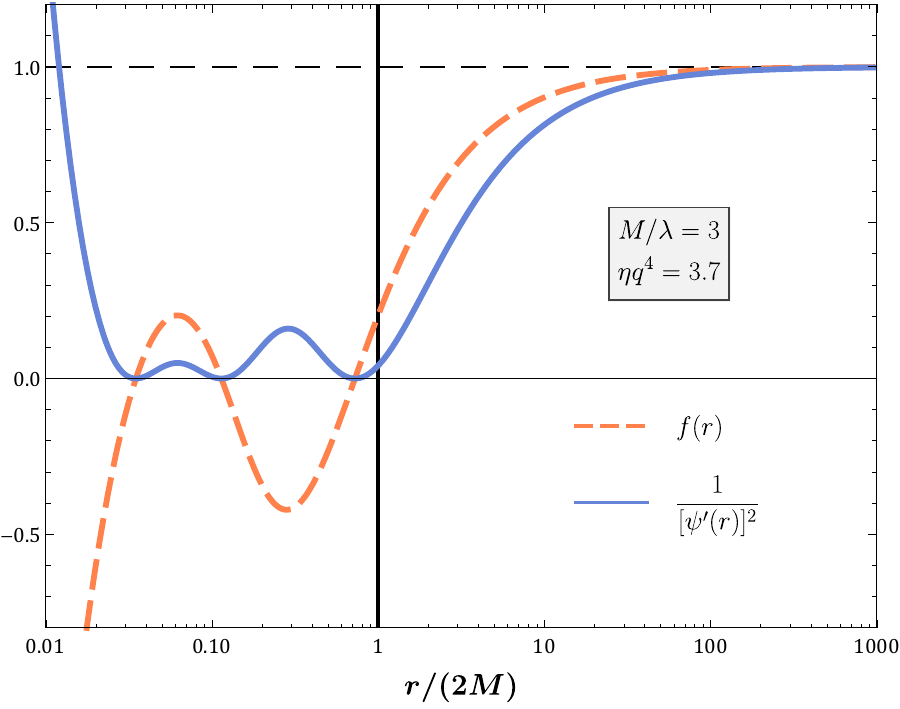}
    \caption{\hspace*{0em}}
    \label{subf: com-phi1}
    \end{subfigure}
    \hfill
    \begin{subfigure}[b]{0.48\textwidth}
    \includegraphics[width=1\textwidth]{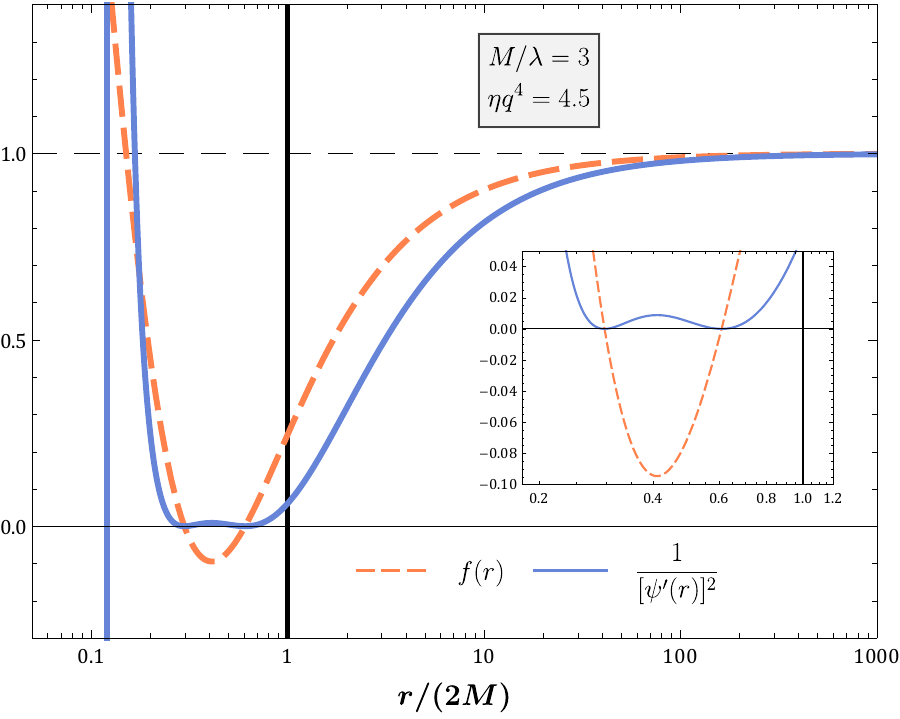}
    \caption{\hspace*{0em}}
    \label{subf: com-phi2}
    \end{subfigure}
    \caption{(a) A three-horizon BH solution with a real-valued scalar field over the entire spacetime. (b) A two-horizon BH solution with a real-valued scalar field in the causal regime $r>r_h$. 
    The horizontal axis is logarithmic in both figures.}
    \label{fig: phi-plots1}
\end{figure}

Our solutions are characterized by a non-trivial scalar field described by \eqref{eq:phi1}. Its radial part is determined through the form of $[\psi'\left(r\right)]^2$. 
Therefore, for a real-valued scalar field, we need the latter quantity to be positive. A sufficient condition for this is $f\left(r\right)\leq 1$, for all $r$. It is not difficult to show that this is always true for $M/\lambda>\pi\eta q^4/4$ or in the case of the regular solutions. In Fig. \ref{subf: com-phi1}, we depict the metric function $f(r)$ for a three-horizon black-hole solution arising in the case when $\eta >0$ and $M/\lambda>\pi\eta q^4/4$ together with the form of $1/[\psi'(r)]^2$. We observe that $[\psi'(r)]^2$ remains always positive thus ensuring a real-valued scalar field over the entire spacetime. The only case in which the scalar field becomes imaginary is when $\eta>0$ and $M/\lambda<\pi\eta q^4/4$, since in this case $f(r) \to +\infty$ as $r\to 0$. However, when this spacetime has a horizon $r_h$, it suffices to ensure that the scalar field remains real for $r>r_h$, and this turns out to be the case. As an illustration, in Fig. \ref{subf: com-phi2}, we display a double horizon black-hole solution where $f(r)$ diverges to infinity as $r \to 0$. Although $[\psi'(r)]^2$ does become negative in the same regime, the scalar field remains real-valued in the entire causal region $r>r_h$.

Let us also briefly comment on the regularity of the scalar field. When the spacetime is a black hole, regularity of $\phi$ at (any) horizon $r_h$ is identified by going to future horizon-crossing coordinates $\left(v,r,\theta,\varphi\right)$ or past horizon-crossing coordinates $\left(u,r,\theta,\varphi\right)$, where $v = t+r^\star$, $u=t-r^\star$ and $r^\star = \int\mathrm{d}r/f\left(r\right)$ are respectively the advanced time, retarded time and tortoise coordinate. Solving (\ref{eq:phi1}) for $\psi'\left(r\right)$, one can choose either sign, $\psi'\left(r\right)=+q/f\left(r\right)\cdots$ or $\psi'\left(r\right)=-q/f\left(r\right)\cdots$. With the $+$ sign, the whole scalar field $\phi$ is regular at any future horizon, since $\phi\left(v,r=r_h\right)=q v + \text{const.}$, and for the $-$ sign, it is regular at any past horizon, $\phi\left(u,r=r_h\right)=q u + \text{const}$. For instance, for the $+$ sign, the scalar field can cross a future event and future inner horizon, but it cannot cross a past inner horizon. One then would have to change coordinates from $v$ to $u$ upon which $\phi $ loses regularity. In other words, the scalar field defines a future pointing direction for the observer, who can hover up to $r=0$ for the regular solution~\eqref{regular} and even go beyond.

In addition, for theories enjoying shift symmetry, it is a common requirement, for example in no-hair theorems~\cite{Hui:2012qt}, that the associated Noether current $J^\mu$ defined in Appendix [see~(\ref{eq:current_app})] have a finite norm in the entire spacetime. As explained in the Appendix, the field equations impose $J^r=0$~\cite{Babichev:2015rva}, so the only non-vanishing component of the current is $J^t$; its magnitude remains finite everywhere apart from the singularity since
\begin{equation}
    J^\mu J_\mu = -f\left(r\right)\left(J^t\right)^2 = -\frac{4\eta^2 q^6\lambda^4\left(11r^2-\lambda^2\right)^2f\left(r\right)}{9\left(r^2+\lambda^2\right)^6}.
\end{equation}

Concerning the theory functionals~(\ref{eq:th1}), they are all well-defined in the whole spacetime being analytic functions of the kinetic term $X$, which is also everywhere finite according to (\ref{eq:X1}). For the stability of the solution~\cite{Takahashi:2019oxz}, we can require in addition that $G_4$ remains positive, meaning that the effective Newton constant is positive, too. This is the case if $\eta q^4<3$. If, on the contrary, $\eta q^4>3$, then $G_4$ is negative below $r=\lambda\sqrt{\sqrt{3\eta}q^2-3}/\sqrt{3}$. Fortunately, the interesting solutions presented above, such as the three-horizon black-hole solutions or the regular black-hole/soliton solutions, emerge also for $\eta q^4<3$. In the case of regular black holes with very large masses, we are forced to have $\eta q^4>3$, but the radius where $G_4$ becomes negative is always hidden behind the horizon.

%
 
\section{Black hole in a theory with canonical kinetic term}\label{sec:bhckt}

A second homogeneous solution with primary hair can be obtained also in a theory that includes a canonical kinetic term, namely:
\begin{equation}
    G_2 = \frac{2\eta}{\lambda^2}X,\quad G_4 = 1+\eta X,\quad F_4 = -\frac{\eta}{4X}.\label{eq:th2}
\end{equation}
As in the previous case, the coupling $\lambda$ has dimension of $\left(\text{length}\right)$ and for simplicity will be assumed positive, while the coupling $\eta$ can take either sign; however, the latter now has dimension $\left(\text{length}\right)^2$. Solving the field equations, we obtain a solution that again admits two integration constants $M$ and $q$ (the primary hair), with metric function given by the expression
\begin{equation}
    f\left(r\right) = 1+\eta q^2-\frac{2M}{r}+\eta q^2\frac{\pi/2-\arctan\left(r/\lambda\right)}{r/\lambda}\,.\label{eq:bh2}
\end{equation}
As regards the scalar field, we obtain  
\begin{equation}
    \phi = qt + \psi\left(r\right),\quad [\psi'\left(r\right)]^2 = \frac{q^2}{f^2\left(r\right)}\left[1-\frac{f\left(r\right)}{1+\left(r/\lambda\right)^2}\right].
\end{equation}
Note that $\psi\left(r\right)$ assumes the same expression in terms of $f\left(r\right)$ as before [see Eq.~(\ref{eq:phi1})], however its explicit dependence on $r$ is now different since $f\left(r\right)$ has changed. One may easily check that the kinetic term $X$ has the exact same expression as in (\ref{eq:X1}). 

The form of the solution seen by an asymptotic observer at infinity follows by taking again the limit $r\to\infty$. This is found to have the form
\begin{equation}
    f\left(r\right) = 1+\eta q^2-\frac{2M}{r}+\eta q^2\frac{\lambda^2}{r^2}+\mathcal{O}\left(\frac{1}{r^4}\right).
\end{equation}
As in the previous case, the integration constant $M$ can be interpreted as the ADM mass of the solution while $q$ appears again in the coefficient of the Reissner-Nordstr\"om-type term. The striking feature in this solution, however, is that the primary hair $q$ introduces also a solid angle excess/deficit---depending on the sign of $\eta$---similar to that of a gravitational monopole \cite{Barriola:1989hx, Chatzifotis:2022ubq}. Indeed, rescaling $r$, as $r\rightarrow \infty$, we obtain
\begin{equation*}
    \mathrm{d}s^2=-\,\mathrm{d}t^2+\mathrm{d}r^2+r^2(1+\eta q^2)\mathrm{d}\Omega^2\,,
\end{equation*}
with the sphere of radius $r$ having now an area equal to $4\pi r^2(1+\eta q^2)$. 
Therefore, the solution is locally asymptotically flat.  If we consider the equator plane at $\theta=\frac{\pi}{2}$, the effect is the same as that of a cosmic string with a conical excess/deficit $-\pi\eta q^2$. The gravitational lensing caused by such defects is well known \cite{Barriola:1989hx}. If the source $S$, black hole $B$, and observer $O$ are perfectly aligned, then the source observed will have the form of a ring, whereas a slight misalignment gives a double image \cite{Barriola:1989hx}.
Consequently, we expect $|\eta q^2|\ll 1$.

\begin{figure}[t!]
    \centering
    \begin{subfigure}[b]{0.49\textwidth}
    \includegraphics[width=1\textwidth]{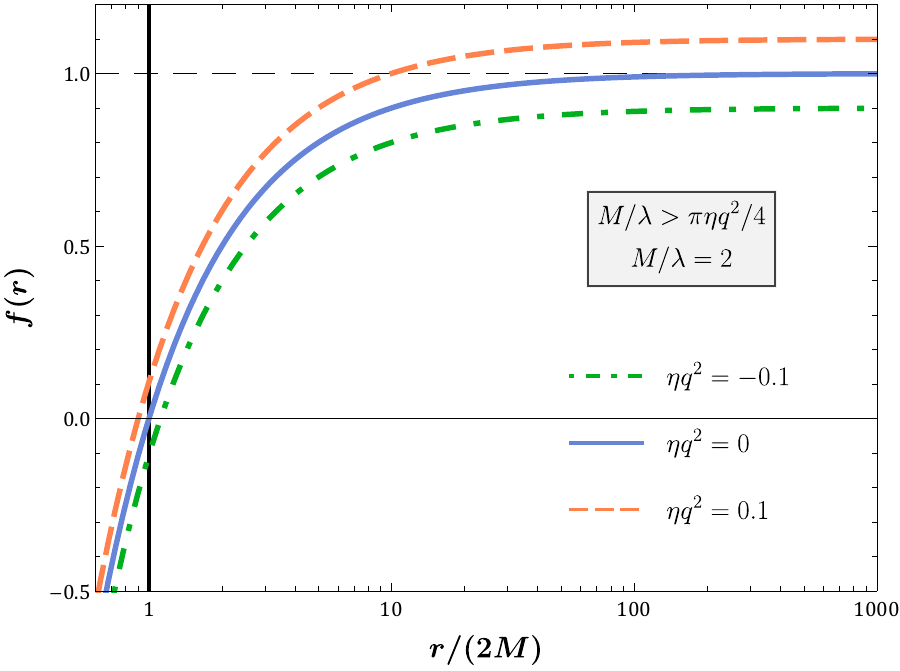}
    \caption{\hspace*{-1.5em}}
    \label{subf: sol2-1}
    \end{subfigure}
    \hfill
    \begin{subfigure}[b]{0.465\textwidth}
    \includegraphics[width=1\textwidth]{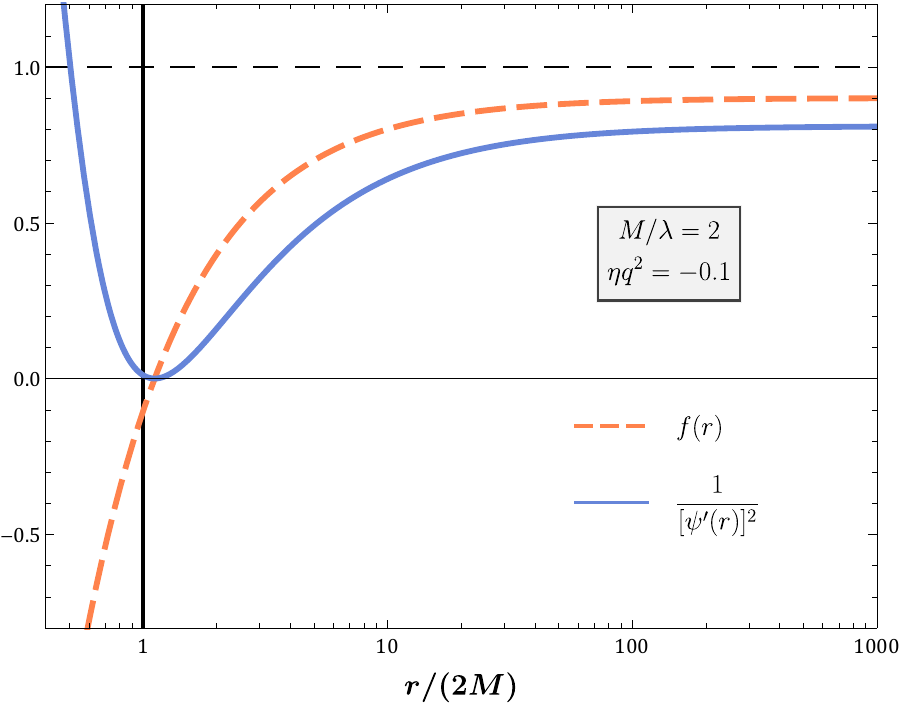}
    \caption{\hspace*{0em}}
    \label{subf: sol2-2}
    \end{subfigure}
    \caption{(a) From bottom to top: a BH solution with angle deficit, the Schwarzschild solution, and a BH solution with angle excess. (b) An indicative BH solution with a real-valued scalar field. The horizontal axis is logarithmic in both figures.}
    \label{fig: sol2}
\end{figure}

We then turn to the radial regime close to the origin. By expanding the solution \eqref{eq:bh2} around $r=0$, we obtain
\begin{equation}
    f\left(r\right) = 1-\frac{2M-\pi\eta q^2\lambda/2}{r}+\frac{\eta q^2r^2}{3\lambda^2}+\mathcal{O}\left(r^4\right).
\end{equation}
As in the previous case, for $\eta<0$, $f\left(r\right)\to -\infty$ when $r\to 0$, therefore the solution describes a black hole with a unique event horizon. Notably, its size is always greater than the Schwarzschild radius, $r_h\geq 2M$, and an increasing function with respect to the scalar charge $q$. The same behaviour was also observed for $\eta <0$ in the case of the asymptotically flat black hole presented in the previous section. Following \cite{Charmousis:2021npl}, it is in fact possible to constrain the scalar hair given that we can probe atomic nuclei of extremely small mass\,\footnote{Strictly speaking, this argument assumes that any static and spherically-symmetric object of the theory~(\ref{eq:th2}) is described by the metric~(\ref{eq:bh2}).} and radius $R\sim 10^{-15}\,$m which are horizonless, i.e. $r_h<R$. An estimation for $r=r_h$ then gives $(-\eta) q^2< \frac{2}{\pi \lambda }\times\left(10^{-15}\,\text{m}\right)$ which constrains the scalar charge for a given coupling constant $\lambda$.
On the other hand, for $\eta>0$, there are again three distinct cases depending on the relative strength of the scalar hair against the mass $M$ of the compact object: for $M/\lambda>\pi\eta q^2/4$, we obtain a black hole with a single event horizon, for $M<\pi\eta q^2\lambda/4$, the spacetime is a naked singularity whereas, for $M/\lambda=\pi\eta q^2/4$, we find a solution with infinitely regular curvature invariants. In the latter case, the solution reads
\begin{gather}
    f\left(r\right) = 1+\frac{4M}{\pi\lambda}\left(1-\frac{\arctan\left(r/\lambda\right)}{r/\lambda}\right),\label{eq:reg_2}\\[2mm]
    \phi = \left(\frac{4M}{\pi\eta\lambda}\right)^{1/2}t + \psi\left(r\right),\quad [\psi'\left(r\right)]^2 = \frac{4M}{\pi\eta\lambda}\frac{1}{f^2\left(r\right)}\left[1-\frac{f\left(r\right)}{1+\left(r/\lambda\right)^2}\right].
\end{gather}
The term between parentheses in (\ref{eq:reg_2}) is always positive, therefore $f\left(r\right)\geq 1$: this time, there is no horizon and the regular spacetime always describes a regular soliton. We note once again that the spacetime (\ref{eq:bh2}) reduces to Schwarzschild for $q=0$, and to flat spacetime for $q=0$ and $M=0$. If $M=0$ but $q\neq 0$, the spacetime is nontrivial despite having no mass: it is a black hole if $\eta<0$ and a naked singularity if $\eta>0$.

In Fig. \ref{subf: sol2-1}, we display an indicative set of black-hole solutions arising in the case when $M/\lambda>\pi\eta q^2/4$ and for positive or negative $\eta$; the two solutions possess a single event horizon with radius smaller or larger than the Schwarzschild radius, respectively. The Schwarzschild solution with $q=0$ is also shown for the sake of comparison. In Fig. \ref{subf: sol2-2}, we depict a black-hole solution arising for $\eta <0$ together with the quantity $1/[\psi'(r)]^2$. The latter remains everywhere positive thus ensuring the reality of the scalar field. 

The regularity of the scalar field follows from the same reasoning as before, by going to horizon-crossing coordinates. The norm of the Noether current is also well-defined everywhere apart from the singularity, with
\begin{equation}
    J^\mu J_\mu = -\frac{\eta^2 q^2\left(3r^2+\lambda^2\right)^2f\left(r\right)}{\left(r^2+\lambda^2\right)^4}.
\end{equation}
Regarding finally $G_4$, and thus the effective gravitational constant, we readily see that it can become negative only for negative values of the coupling $\eta$, namely for $\eta q^2<-2$. In this case, $G_4$ is negative below $r=\lambda\sqrt{-1-\eta q^2/2}$. However, it turns out that this radius is again always hidden behind the horizon.

%

\section{Conclusions and discussion}

In this work, we have found for the first time in the literature explicit, simple examples of asymptotically flat black holes with primary scalar hair described by the  metrics~(\ref{eq:bh1}) and~(\ref{eq:bh2}). The theories where these solutions emerge are (single) scalar-tensor theories, and more particularly, simple beyond Horndeski theories given respectively by~(\ref{eq:th1}) and~(\ref{eq:th2}). When the scalar charge is set to zero, we recover the Schwarzschild solution but, unlike other numerous numerical or explicit solutions, the scalar hair is a free integration constant. This is in contrast to theories which exhibit scalarisation where the scalar hair is fixed with respect to the mass in order to have a regular solution at the horizon. Here, the scalar hair $q$ of the solutions is associated to the linear time-dependence in the scalar field, and hence to the shift symmetry in the action. These types of solutions are known to be regular at the horizon precisely as a result of the linear time dependence, and therefore no constraint of regularity is necessary on the horizon. Quite contrary to previous solutions with a time-dependent scalar field, the kinetic term here is not constant, and the primary hair appears in the spacetime metric. Hence, our solutions cannot be classified as stealth solutions. 

The black-hole solutions we have determined  include geometries with one, two or three horizons depending on the values of the parameters $M$, $q$, and the coupling constants of the theory. 
The solutions emerging can have an event horizon radius either larger or smaller than the Schwarzschild radius, and thus they can be more sparse or more compact compared to the GR solution. In all cases, the kinetic term of the scalar field is everywhere well-defined.

A remarkable feature of our solutions is that, for a particular relation between the mass and the scalar hair, they become regular spacetimes (solitons or black holes), i.e. their curvature invariants are infinitely regular for all values of the coordinates. These regular metrics are given by Eqs.~(\ref{eq:reg}) and~(\ref{eq:reg_2}) while the scalar is regular throughout spacetime. Regular black holes have already been found to exist in DHOST theories \cite{Babichev:2020qpr, Baake:2021jzv}. Here however, it is the first time that the solutions are not catered to a specific theory, they are obtained for a particular relation in-between the black hole parameters within a generic black hole solution.

It would be very important to study perturbations of the solutions at hand -- there, one may expect that excessive scalar hair renders these solutions unstable so as to evade naked singularities, which emerge for large values of $q$. Also, one would like to examine if there are precise examples of spontaneous scalarisation, in the sense that solutions with $q\neq 0$ are preferred compared to $q=0$ solutions. Unlike spontaneous scalarisation where one has two (at least) branches of solutions, here we have a single solution with two free parameters. As such, one can continuously go from the solution with $q \neq 0$ to the solution with no hair. 

How generic are the solutions that we have found? First, one can check that if we impose a canonical kinetic term $G_2\propto X$ and a homogeneous solution, $f=h$, the formalism developed in the Appendix unambiguously leads to the primary hair solution discussed in Sec.~\ref{sec:bhckt}. Second, one can question the necessity of going beyond Horndeski to get such primary hair solutions. By setting $F_4=0$ and $f=h$ in the field equations presented in the Appendix, one is led to solutions with unsatisfying features, typically, a solid angle deficit which depends on the coupling constants [and not only on the hair $q$ as in the solution~(\ref{eq:bh2})], or non-physical asymptotic terms similar to those described in~\cite{Babichev:2017guv}. We can however identify one interesting case in this pure Horndeski framework, which however has little to do with a primary hair black hole, and rather appears as a byproduct of the formalism developed in the Appendix. The theory is 
\begin{equation}
    G_2 = -2\Lambda+2\eta\sqrt{X},\quad G_4 = 1+\lambda\sqrt{X},\quad F_4=0,
\end{equation}
with two coupling constants $\eta$ and $\lambda$, and a cosmological constant $\Lambda$. The metric is nothing but Schwarzschild-de Sitter (or Schwarzschild if $\Lambda=0$),
\begin{equation}
    f\left(r\right) = 1-\frac{2M}{r}-\frac{\Lambda r^2}{3},
\end{equation}
while the kinetic term and scalar field are
\begin{equation}
    X = \frac{\lambda q^2/2}{\lambda+\eta r^2},\quad \phi(t,r) = qt + \psi\left(r\right),\quad [\psi'\left(r\right)]^2 = \frac{q^2}{f^2\left(r\right)}\left[1-\frac{\lambda f\left(r\right)}{\lambda+\eta r^2}\right].
\end{equation}
This is therefore a stealth solution, but, as opposed to all previously described stealth solutions~\cite{Babichev:2013cya,Kobayashi:2014eva}, the scalar field has a non-constant kinetic term. $q$ is an arbitrary integration constant of the solution, but should not be thought of as primary hair, since it does not appear in the metric. While usual stealth solutions, with constant $X$, are prone to strong coupling issues according to~\cite{Babichev:2018uiw,deRham:2019gha}, it is surely worth investigating if this new stealth solution enjoys a healthier behaviour. It would also be interesting to investigate the existence of de Sitter solutions with primary hair and see how these are related (or not) to the self-tuning de Sitter stealth solutions with a constant kinetic term \cite{Babichev:2013cya,Charmousis:2015aya}.

As another interesting perspective, one may wonder if one could construct in a similar way inhomogeneous solutions that allow for black holes and wormhole metrics at the same time. Is it possible that the scalar charge in this case allows for the appearance of a wormhole throat instead of a black hole horizon? We know that beyond Horndeski theories allow for such geometries \cite{Bakopoulos:2021liw,Chatzifotis:2021hpg,Babichev:2022awg}, so a similar construction may be possible here. One can also try to extend this construction to theories without parity or shift symmetry, and figure out if these theories originate from a higher dimensional metric theory such as Lovelock theory where no apparent symmetries are present \cite{Babichev:2013cya,Babichev:2023psy}. These are some of the exciting questions one could try to answer in the near future. 

\acknowledgments
We would like to thank Eugeny Babichev, Nikos Chatzifotis, Mokhtar Hassaine and Karim Noui  for useful discussions. The research project was supported by the Hellenic Foundation for Research and Innovation (H.F.R.I.) under the ``3rd Call for H.F.R.I. Research Projects to support Post-Doctoral Researchers" (Project Number: 7212). This work was supported by the French National Research Agency via Grant No. ANR-20-CE47-0001 associated with the project COSQUA (Cosmology and Quantum Simulation).
NL acknowledges support of ANR grant
StronG (ANR-22-CE31-0015-01) and of the
doctoral program Contrat Doctoral Sp\'ecifique Normalien \'Ecole
Normale Sup\'erieure de Lyon (CDSN ENS Lyon).

\appendix\label{appa}

\section{Field Equations}
This Appendix explains how, among the general shift and parity-symmetric beyond Horndeski theories, see Eq.~(\ref{eq:action}), we identified the two theories presented above, Eqs.~(\ref{eq:th1}) and~(\ref{eq:th2}), as admitting black-hole solutions. The setup is therefore the action $S$ of Eq.~(\ref{eq:action}), with static, spherically-symmetric ansatz~(\ref{eq:metric_ansatz}) for the metric, and ansatz~(\ref{eq:scalar_ansatz}) for the scalar field, with a linear time dependence. The kinetic term is $X=-\frac{1}{2}\partial^\mu\phi\partial_\mu\phi$, thus $2X=q^2/h-f\psi'^2$. Because of shift-symmetry, there exists a Noether current,
\begin{equation}
    J^\mu = \frac{1}{\sqrt{-g}}\frac{\delta S}{\delta\left(\partial_\mu\phi\right)}.\label{eq:current_app}
\end{equation}
In general, the scalar field equation then reads $\nabla_\mu J^\mu=0$, but it was shown in~\cite{Babichev:2015rva} that, with the considered ansatz~(\ref{eq:metric_ansatz}) for the metric and~(\ref{eq:scalar_ansatz}) for the scalar field, the independent field equations are the Noether current component $J^r=0$, and the metric field equations $tt$ and $rr$, which results in three equations for three unknowns $h\left(r\right)$, $f\left(r\right)$ and $\psi\left(r\right)$. We denote these three independent equations as $\mathcal{E}_J$, $\mathcal{E}_t$ and $\mathcal{E}_r$. Following~\cite{Bakopoulos:2022csr}, we introduce the useful quantity
\begin{align}
Z\left(X\right)={}&{}2XG_{4X}-G_4+4X^2F_4,\label{eq:zdef}
\end{align}
which essentially replaces $F_4$ and enables to write the field equations as
\begin{align}
\mathcal{E}_J={}& -\frac{2fh'}{h}rZ_X+r^2G_{2X}+2G_{4X}-2fZ_X+\frac{q^2f}{X h}\left(Z_X-G_{4X}\right)+\frac{2q^2f}{h}rF_4\left(\frac{f'}{f}-\frac{h'}{h}\right),\\
\mathcal{E}_r={}& -\frac{2fh'}{h}rZ-r^2G_2-2G_4-2fZ+\frac{q^2f}{X h}\left(Z+G_4 \right)-\frac{4q^2f}{h}rF_4X'-f\left(\psi'\right)^2\mathcal{E}_J,\\
\mathcal{E}_t={}& 4rX'Z_X+2\left(\frac{f'}{f}-\frac{h'}{h}\right)rZ-\frac{\mathcal{E}_r}{f}-\left(\left(\psi'\right)^2+\frac{q^2}{fh}\right)\mathcal{E}_J.
\end{align}
The complete system of field equations, $\mathcal{E}_J=0$, $\mathcal{E}_r=0$ and $\mathcal{E}_t=0$, thus gives the following system
\begin{align}
2X'Z_X ={}&{} \left(\frac{h'}{h}-\frac{f'}{f}\right)Z, \label{eq1}\\
\frac{2fh'}{h}rZ_X = {}&{} r^2G_{2X}+2G_{4X}-2fZ_X+\frac{q^2f}{X h}\left(Z_X-G_{4X}\right)+\frac{2q^2f}{h}rF_4\left(\frac{f'}{f}-\frac{h'}{h}\right),\label{eq2}\\
\frac{2fh'}{h}rZ={}&{}-r^2G_2-2G_4-2fZ+\frac{q^2f}{X h}\left(Z+G_4 \right)-\frac{4q^2f}{h}rF_4X'\label{eq3}.
\end{align}
This system can be simplified further. First of all, (\ref{eq1}) integrates to
\begin{equation}
\label{easy}
 \frac{f}{h}=\frac{\gamma^2}{Z^2},
\end{equation}
with $\gamma$ a constant. Also, (\ref{eq2}) and~(\ref{eq3}) combine to give
\begin{equation}
\label{Xeq}
    r^2(G_2 Z)_{X}+2(G_4 Z)_X \left(1-\frac{q^2 \gamma^2}{2 Z^2 X}\right)=0,
\end{equation}
while~(\ref{eq3}) combined with~(\ref{easy}) yields 
\begin{equation}
\label{heq}
    2\gamma^2 \left(h r-\frac{q^2 r}{2 X}\right)'=-r^2 G_2 Z-2G_4 Z\left(1-\frac{q^2 \gamma^2}{2 Z^2 X}\right)+\frac{q^2\gamma^2X' r}{ZX^2}\left( 2XG_{4X}-G_4\right).
\end{equation}
We now briefly review the method described in~\cite{Bakopoulos:2022csr}, with the use of the auxiliary function $\G$. Indeed, consider some function $\G$ such that 
\begin{equation}
    \G_X=\frac{\alpha r^2+\tilde\beta(X)}{\epsilon r^2+\tilde\delta(X)},
\end{equation}
where $\alpha$, $\epsilon$ are constants while $\tilde \beta(X)$, $\tilde \delta(X)$ are functions which will be fixed shortly. Indeed, to ensure compatibility of $\G$ and (\ref{Xeq}) we need to have
\begin{align}
    G_2 Z={}&{}\epsilon \G -\alpha X+ C,\\
    2G_4 Z={}&{}\delta \G-\beta X+ D,
\end{align}
while $\tilde \beta(X)=\beta \left(1-\frac{q^2 \gamma^2}{2 Z^2 X}\right)$, $\tilde \delta(X)=\delta \left(1-\frac{q^2 \gamma^2}{2 Z^2 X}\right)$, and $\beta$, $\delta$, $C$ and $D$ are constants. As a consequence, once $\G$ is given explicitly as a function of $X$ and $Z$ is fixed, one knows immediately the theory functionals $G_2$ and $G_4$. We note that choosing $Z$ amounts to fixing the relation between $f$ and $h$, see~(\ref{easy}), as well as $F_4$, see~(\ref{eq:zdef}). Then, by using (\ref{Xeq}), $X$ is found algebraically as a function of $r$. Finally, finding $h$ (and thus $f$) is a matter of direct integration of (\ref{heq}). 

Let us consider the case studied for $q=0$ in~\cite{Bakopoulos:2022csr}, namely $Z=\gamma$ (i.e. $h=f$) and $\G_X=2\mu X+\zeta$ with $\mu$ and $\zeta$ constants. Then,
\begin{align}
    G_2={}&{}\frac{\epsilon \mu}{\gamma}X^2+\frac{\epsilon \zeta-\alpha}{\gamma}X-2\Lambda,\\
    G_4={}&{}\frac{\delta \mu}{2\gamma}X^2+\frac{\delta \zeta-\beta}{2\gamma}X+1,
\end{align}
where we have fixed the constants $C$ in $G_2$ to give the usual bare cosmological constant $\Lambda$ and $D$ in $G_4$ to give a usual Einstein-Hilbert term. The definition of $Z$ leads to the following expression for the beyond Horndeski potential $F_4$,
\begin{equation}
    F_4 = \frac{\gamma+1}{4X^2}+\frac{\beta-\delta\zeta}{8\gamma X}-\frac{3\delta\mu}{8\gamma}.
\end{equation}
This directly translates (\ref{Xeq}) to an algebraic equation for $X$,
\begin{equation}
    2\mu (\epsilon r^2+\delta)X^2+\left[(\zeta\epsilon -\alpha)r^2+\delta \zeta-\beta-q^2\mu\delta\right]X-(\delta\zeta-\beta)\frac{q^2}{2}=0. \label{xpoly}
\end{equation}
In general, the solution for $X$ displays square roots, and Eq.~(\ref{heq}) for $h$ can be integrated only formally, giving an expression for $h$ depending on a complicated integral. However, when the equation reduces to a linear equation for $X$, then~(\ref{heq}) is integrated explicitly. This corresponds to the two cases studied in the article: by removing the $X^0$ term, i.e. $\delta\zeta-\beta=0$,\,\footnote{One must also choose $\alpha-\epsilon\zeta=0$ to achieve asymptotic flatness in this case.} one gets the case studied in Sec.~\ref{sec:afbh}, while by removing the $X^2$ term, i.e. $\mu=0$, one gets the case studied in Sec.~\ref{sec:bhckt}. In particular, we focused on the case $\Lambda=0$ but the presented solutions have an immediate de Sitter generalization for nonzero $\Lambda$. In addition, we appropriately chose the values of some of the coupling constants to get relevant asymptotics (for example, it turns out that one must end setting $\gamma=-1$). Finally, we have renamed all coupling constants in order to obtain the simple expressions~(\ref{eq:th1}) and~(\ref{eq:th2}) for the theories and for their corresponding solutions.

Additionally, although $h$ has no closed form expression when~(\ref{xpoly}) is a true quadratic equation for $X$, it remains possible to study explicitly the asymptotics of the metric, which in every case displays a primary hair $q$. 
Interestingly, the only case which leads to a pure asymptotically flat black hole for $\Lambda=0$ as in Sec.~\ref{sec:afbh}, that is $f\left(r\right)=1-2M/r+\mathcal{O}\left(1/r\right)$ at infinity, is precisely the case of Sec.~\ref{sec:afbh}. It remains possible to get $f\left(r\right)=1-2M/r+\mathcal{O}\left(1/r\right)$ for other theories, but at the price of introducing a nonvanishing bare cosmological constant $\Lambda$ whose value must be adjusted with respect to the other coupling constants.
On the other hand, several other theories lead to a black hole with similar asymptotics as in Sec.~\ref{sec:bhckt}, that is with a solid angle deficit triggered by the primary hair $q$. Among those, we chose to focus on the theory of Sec.~\ref{sec:bhckt}, because it offers a closed form expression and is the only one where $G_2\propto X$ is a canonical kinetic term.

\addcontentsline{toc}{section}{References}
\bibliography{Refs}{}
\bibliographystyle{utphys}

\end{document}